\documentclass[prd,nofootinbib,onecolumn]{revtex4}

\usepackage{amssymb}
\usepackage{cases}
\usepackage{graphicx}
\usepackage{latexsym}

\newcommand{\hmpc}{{\rm \,h/Mpc}}
\newcommand{\Cl}{C_\ell}

\begin{document}

\title{Constraints on Massive Neutrinos from the CFHTLS Angular Power Spectrum}
\author{Jun-Qing Xia${}^1$}
\author{Benjamin R. Granett${}^2$}
\author{Matteo Viel${}^{3,4}$}
\author{Simeon Bird${}^{5}$}
\author{Luigi Guzzo${}^{2}$}
\author{Martin G. Haehnelt${}^{6}$}
\author{Jean Coupon${}^{7}$}
\author{Henry Joy McCracken${}^{8}$}
\author{Yannick Mellier${}^{8}$}

\affiliation{${}^1$Scuola Internazionale Superiore di Studi
Avanzati, Via Bonomea 265, I-34136 Trieste, Italy}

\affiliation{${}^2$INAF-Osservatorio Astronomico di Brera, Via E.
Bianchi 46, 23807, Italy}

\affiliation{${}^3$INAF-Osservatorio Astronomico di Trieste, Via
G.B. Tiepolo 11, I-34131 Trieste, Italy}

\affiliation{${}^4$INFN/National Institute for Nuclear Physics, Via
Valerio 2, I-34127 Trieste, Italy}

\affiliation{${}^5$School of Natural Sciences, Institute for
Advanced Study, Princeton, NJ 08540, USA}

\affiliation{${}^6$Institute of Astronomy and Kavli Institute for
Cosmology, Madingley Road, CB3 0HA, Cambridge, United Kingdom}

\affiliation{${}^7$Institute of Astronomy and Astrophysics, Academia
Sinica, P.O. Box 23-141, Taipei 10617, Taiwan}

\affiliation{${}^8$Institut d'Astrophysique de Paris, UMR 7095 CNRS,
Universit\`{e} Pierre et Marie Curie, 98 bis Boulevard Arago, 75014
Paris, France}

\date{\today}

\begin{abstract}
We use the galaxy angular power spectrum at $z\sim0.5-1.2$ from the
Canada-France-Hawaii-Telescope Legacy Survey Wide fields
(CFHTLS-Wide) to constrain separately the total neutrino mass
$\sum{m_\nu}$ and the effective number of neutrino species
$N_{\rm{eff}}$. This survey has recently benefited from an accurate
calibration of the redshift distribution, allowing new measurements
of the (non-linear) matter power spectrum in a unique range of
scales and redshifts sensitive to neutrino free streaming. Our
analysis makes use of a recent model for the effect of neutrinos on
the weakly non-linear matter power spectrum derived from accurate
N-body simulations. We show that CFHTLS, combined with WMAP7 and a
prior on the Hubble constant provides an upper limit of
$\sum{m_\nu}<0.29\,$eV and $N_{\rm{eff}} =4.17^{+1.62}_{-1.26}$
(2$\,\sigma$ confidence levels). If we omit smaller scales which may
be affected by non-linearities, these constraints become
$\sum{m_\nu}<0.41\,$eV and $N_{\rm{eff}} =3.98^{+2.02}_{-1.20}$
(2$\,\sigma$ confidence levels). Finally we show that the addition
of other large scale structures probes can further improve these
constraints, demonstrating that high redshift large volumes surveys
such as CFHTLS are complementary to other cosmological probes of the
neutrino mass.
\end{abstract}

\maketitle

\section{Introduction}\label{int}

Determining the neutrino mass is one of the great unsolved problems
of modern particle physics. The standard model contains three
massless neutrino species; observations of neutrino oscillations in
atmospheric and solar neutrino experiments have confirmed that
neutrinos are massive, but cannot pin down their absolute mass
scale. Fortunately, the Universe offers a new laboratory for
neutrino physics. Massive neutrinos affect both the background
expansion and the growth of cosmological structure, making
cosmological observations an unrivaled probe of the total neutrino
mass \cite{NeuRev1,NeuRev2}.

The redshift at which massive neutrinos become non-relativistic
alters the time since matter-radiation equality, thus changing the
position of the Cosmic Microwave Background (CMB) anisotropy peaks.
Measurements of the CMB have used this effect to find
$\sum{m_\nu}<1.3\,$eV (95\% C.L.)
\cite{Spergel2003,Spergel2007,Komatsu2009,Komatsu2011}, and a
sensitivity of $\sigma(\sum{m_\nu}) \sim 0.3\,$eV could soon be
achieved with the Planck satellite (e.g. \cite{NeuRev2}).

Massive neutrinos also play a role in the formation of large scale
structure. Once non-relativistic, they damp the growth of
perturbations within their free streaming scale, resulting in a
suppression of the small-scale linear matter power spectrum of
$\Delta P/P\sim-8\Omega_\nu/\Omega_m$ \cite{Hu1998}. Measurements of
the matter power spectrum can thus improve constraints on the
neutrino mass considerably. Many analyses have been performed
combining CMB data with Large Scale Structure (LSS) probes such as
the 2dF Galaxy Redshift Survey (2dFGRS)
\cite{Elgaroy2002,Allen2003,Barger2004}, the Sloan Digital Sky
Survey (SDSS) \cite{Hannestad2003,Tegmark2004,Hannestad2004,
  Crotty2004,Elgaroy2005,Hannestad2005,Fogli2006,Xia2008,Reid2010a,
  Reid2010b,Reid2010c,Thomas2010,Sekiguchi2010,Saito2011}, WiggleZ
\cite{Riemer2011} and the SDSS Lyman-$\alpha$ forest
\cite{Seljak2005,Goobar2006,Seljak2006} to constrain $\sum{m_\nu}$.
The combination of the SDSS DR8 LRG angular power spectra, WMAP7
data and an HST prior on the Hubble constant gives
$\sum{m_\nu}<0.26\,$eV (95\% C.L.), assuming a flat $\Lambda$CDM
model with the standard model effective number of neutrino
species, $N_{\rm eff}=3.04$ \cite{Hoetal2012,DePutter2012}.

A detection of $N_{\rm eff} > 3.04$, as recently hinted at by the
Atacama Cosmology Telescope (ACT) and the South Pole Telescope (SPT)
experiments \cite{Dunkley2011,Keisler2011}, would imply additional
relativistic relics or non-standard neutrino properties. The
additional energy density of extra relativistic particle species
would change the redshift of matter-radiation equality, leaving
imprints on the CMB anisotropies and matter power spectrum.
Parameter degeneracies limit the constraining power of the primary
CMB anisotropies on $N_{\rm eff}$
\cite{Spergel2007,Dunkley2009,Komatsu2009}. The 95\% lower limit is
$N_{\rm eff}>2.7$ from WMAP7 alone \cite{Komatsu2011}. Adding
information on the matter power spectrum and the Hubble constant can
tighten the constraint significantly
\cite{Ichikawa2007,Mangano2007,Hamann2007,Mantz2010,Reid2010a,Reid2010c,Keisler2011},
as can measurements of smaller scale CMB fluctuations.

In this work, we introduce new constraints from measurements of the
angular power spectrum of galaxies at $z\sim 0.5-1.2$ in the
Canada-France-Hawaii-Telescope Legacy Survey Wide fields
(CFHTLS-Wide) \cite{CFHTLS}. This is the deepest wide-field survey
of its kind, covering 133 sqr. deg. and sampling a comoving volume
of 0.2 Gpc$^3/$h$^3$ from $z=0.5-1.2$. This data has been used for
cosmological studies, including weak lensing constraints
\cite{Fu2008,Li2009,Kilbinger2009,Shan2012}.  In particular, a joint
analysis using the weak lensing measurements has given a limit for
massive neutrinos of $\sum{m_\nu}<0.54\,$eV (95\% C.L.)
\cite{Tereno2009}.  The small-scale clustering was studied using the
halo model by ref. \cite{Coupon012} and the deprojected power
spectrum was presented in ref. \cite{Granett2012}.

Considerable gains can be made by probing the power spectrum on
weakly non-linear scales, provided one can model the galaxy bias
sufficiently accurately. On large scales, the galaxy power spectrum
is well-fit by the dark matter power spectrum and a linear galaxy
bias: $\delta_{g}=b_g\delta_{dm}$. On small scales, where the
details of galaxy formation come into play, this relation breaks
down and a general, scale-dependent, galaxy bias model is necessary
\cite{Percival2007,Smith07,Cresswell2009}. This results in a
degeneracy between galaxy bias and cosmological parameters that is
difficult to break without further observations such as
gravitational lensing \cite{Schneider1998}.

The scale-dependence of galaxy clustering has been investigated with
simulations \cite{Smith07,Hamaus2010} and observations
\cite{Cresswell2009}. These studies show that galaxies with lower
luminosity have a weaker scale-dependence of the bias than the most
luminous ones at $k>0.1\hmpc$.  Additionally, the scale dependence
becomes more severe for strongly biased tracers; negligible scale
dependence of the halo bias is found in simulations at $k=0.15\hmpc$
for halos with $b_g<2$, while for more massive halos with $b_g>2$,
there is a 20\% effect \cite{Hamaus2010}.  For the SDSS main sample,
the {\tt Halofit} non-linear matter power spectrum with a constant
bias factor has been demonstrated to be a good fit to the galaxy
power spectrum \cite{Swanson2010}, while the power spectrum of the
LRG sample diverges at $k=0.2\hmpc$ \cite{Percival2007}.

We select the range of scales to study based on the dominance of the
two-halo term in the halo occupation distribution (HOD) model of the
power spectrum.  In ref. \cite{Coupon012}, HOD fits are made to the
correlation function of the CFHTLS galaxies.  We have checked
explicitly that the two-halo term is greater than the one-halo term
to $\ell=960$ in our lowest redshift sample ($0.5<{\rm
z_{phot}}<0.6$). In Sec. \ref{results} we test our CFHTLS data set
using a comparison of the constant bias model with a two-parameter
model in the context of $\Lambda$CDM without massive neutrinos. We
conclude that a constant bias model is sufficient.

The CFHTLS data set is ideal for this study because (i) at higher
redshift the onset of non-linear growth happens at smaller scales; and
(ii) we are targeting ``normal'' galaxies: the flux limited sample
that we construct ($i_{\rm AB}<22.5$) selects galaxies to $z=1$ with
luminosities of $M_g\sim-20$, similar to the SDSS main sample at
$z=0.1$ \cite{Coupon09,Zehavi2011}.

Since we analyze the projected density field, any modulation of the
power spectrum due to redshift-space distortions is minimized,
although these effects must be considered for surveys covering a
larger fraction of the sky than CFHTLS \cite{Thomas2010}. We must
also consider systematic errors arising from the luminosity
dependence of the galaxy bias. In a flux-limited sample with a
mixture of galaxy types, luminosity-dependent biasing can modify the
slope of the galaxy power spectrum.  However, for current surveys
such as 2dFGRS, the effects can be neglected \cite{Percival2004}.

In this paper we will present constraints on the total mass of
neutrinos, $\sum{m_\nu}$, and the effective number of neutrino
species, $N_{\rm eff}$, from different cosmological observations.
The structure of the paper is as follows: In section \ref{cltheory}
we briefly review the general formalism of the angular power
spectrum $C_\ell$. In sections \ref{cldata} and \ref{otherdata} we
describe the recently released CFHTLS clustering power spectra data
set, and the other data sets we use. Section \ref{results} contains
our main results on $\sum{m_\nu}$ and $N_{\rm eff}$, while section
\ref{summary} is dedicated to the conclusions and discussion.

\section{Angular Power Spectrum}\label{cltheory}

The angular power spectrum, $C_\ell$, is a projection of the spatial
power spectrum of fluctuations, $P_{\rm gal}(k,z)$, where $k$ is the
comoving wave number and $z$ is the redshift. The equation for the
projection is
\begin{equation}
C_\ell = \frac{2b^2}{\pi}\int k^2dkP(k)g^2_\ell(k)~,\label{eq:cl}
\end{equation}
where $P(k)$ is the matter power spectrum today and $b$ is the
assumed scale-independent bias factor relating the galaxy
overdensity to the mass overdensity. In our calculations, we assume
the values of bias are constant in each redshift bin; with width
$\Delta z \sim 0.2$, these are relatively narrow. We also neglect
the effect of redshift space distortions, as their effect is limited
to large scales, $\ell < 30$ \cite{Padmanabhan2007}, and our largest
bin is at $\ell = 40$. With these assumptions, the kernel
$g_\ell(k)$ is given by
\begin{equation}
g_\ell(k)=\int^{z_{\rm max}}_{z_{\rm
min}}dzD(z)j_\ell(k\chi(z))\frac{dN}{dz}(z)\left[\frac{dz}{d\chi}(z)\right]~,\label{eq:gl}
\end{equation}
where $j_\ell(x)$ is the spherical Bessel function, $\chi(z)$ is the
comoving distance to redshift $z$, and $dN/dz(z)$ is the normalized
redshift distribution of the survey.

The Limber approximation is valid for all but the largest angular
scales ($\ell \gtrsim 10$). For $\ell\gg1$, we have
\begin{equation}
\frac{2}{\pi}\int
k^2dkj_\ell(k\chi)j_\ell(k\chi')=\frac{1}{\chi^2}\delta(\chi-\chi')~,
\end{equation}
and the equation for the projection becomes
\begin{equation}
C_\ell=b^2\int^{z_{\rm max}}_{z_{\rm
min}}P\left(k=\frac{\ell+1/2}{\chi(z)},z\right)\left[\frac{dN}{dz}(z)\right]^2\frac{dz}{dV_c(z)}~,
\end{equation}
where $dV_c(z)$ is the comoving volume element,
$dV_c(z)=\chi^2d\chi/dz$.

\section{$C_\ell$ Measurements of CFHTLS}\label{cldata}
\newcommand{\nside}{{\rm n_{side}}}
\newcommand{\n}{{\bf \hat{n}}}
\newcommand{\x}{{\bf x}}
\newcommand{\C}{{\bf C}}
\newcommand{\Cinv}{{\bf C}^{-1}}
\newcommand{\Pl}{\mathcal{P}_\ell}
\newcommand{\Tr}{{\rm Tr}}

\subsection{Data}

We analyze the galaxy density maps constructed by ref.
\cite{Granett2012} from the Canada-France-Hawaii Legacy Survey Wide
fields (CFHTLS-Wide). The data set is based on the CFHTLS T0006
release, including photometric redshift estimates \cite{Coupon09}. A
selection of galaxies was made photometrically to an apparent flux
limit of $i_{\rm AB}=22.5$ in three photometric redshift bins:
$0.5-0.6$, $0.6-0.8$, and $0.8-1.0$, labeled S6, S7 and S8. We are
not overly concerned with the error in the photometric redshift
estimates, although it is typically $\Delta z/(1+z)=0.04$
\cite{Coupon09}; we only require knowledge of the redshift
distributions of the samples.  In ref.  \cite{Granett2012}, the
redshift distributions were measured using a subset of $\sim$14000
spectroscopic redshifts obtained from the VIMOS Public Extragalactic
Redshift Survey (VIPERS) \cite{Guzzo2012}.  We also use those
redshift distributions here.

\subsection{$\Cl$ estimator}

We estimate the $\Cl$ with a maximum likelihood approach first
applied to cosmic microwave background maps \cite{Tegmark97,Bond98}.
It is well suited to fields that may be described by a Gaussian
likelihood function, and it has been extended successfully to
measure the angular power spectrum in galaxy surveys
\cite{Huterer01,Tegmark02}.

For surveys that cover only a relatively small patch of sky, such as
CFHTLS, the maximum likelihood approach has advantages over other
techniques.  Generally, it is challenging to measure harmonic modes
that are strongly affected by the survey geometry.  Modes become
correlated due to the convolution effect of the survey mask;
further, the hard edges can introduce `ringing' in harmonic space.
Our estimator accounts for these effects and measures the $\Cl$ with
minimum variance (in the case of a Gaussian field).  Along with the
$\Cl$, we derive the covariance matrix of the measurements and
window functions.

The galaxy density maps were constructed on a {\tt
HEALPix}\footnote{\url{http://healpix.jpl.nasa.gov}} grid with an
angular resolution of 7 arcmin ($\nside=512$) \cite{Healpix}.  We
order the $m$ pixels of the density map, form a data vector,
$\x=[\delta(\n_0),\delta(\n_1),...,\delta(\n_{m-1})]$, and write the
covariance of the data as $C_{ij} = \langle x_i x_j \rangle$.  The
pixel covariance matrix is given by the sum of the signal and the
noise components
\begin{equation}
C_{ij} = \sum_l \frac{2l+1}{4\pi} \Pl(\cos \theta_{ij}) B_l^2 \Cl + N_{ij}\,,
\end{equation}
where $\Pl$ are the Legendre polynomials, $\theta_{ij}$ is the
separation between pixels $i$ and $j$ and $N_{ij}$ is the noise
covariance matrix.  The noise matrix is taken to be diagonal with
Poisson elements given by $N_{ii}=w_i^2/\bar{n}$, where $w_i$ is a
weight accounting for partially-sampled pixels.  The finite
resolution of the pixelised map attenuates the power spectrum by the
pixel window function, $B_\ell$, which depends on the pixel geometry
\cite{Healpix}.

To simplify the following expressions, we introduce a symmetric
matrix ${\bf P}_b$.  This is a sum of the Legendre polynomials over
a band-power indexed by $b$ including $[\ell_b,\ell_{b+1})$.  The
components are
\begin{equation}
P_{b,ij} = \sum_{\ell_b\le\ell<\ell_{b+1}}  \frac{2\ell+1}{4\pi} \Pl(\cos \theta_{ij}) B_\ell^2\,,
\end{equation}

The quadratic band-power estimator is given by
\begin{equation}
{\hat C}_b = \frac{1}{2} \sum_{j}A_{ij}  \left\{  \x^T {\bf C}^{-1} {\bf P}_b {\bf C}^{-1} \x  - \Tr \left( {\bf C}^{-1} {\bf P}_b {\bf C}^{-1} {\bf N}\right)\right\}\,.
\label{eq:est}
\end{equation}
The matrix ${\bf A}$ is a mixing matrix that sets the normalization
and may be specified to form linear combinations of the bin
estimates. It is related to the Fisher matrix,
\begin{equation}
F_{bb'}=\frac{1}{2} \Tr \left( {\bf C}^{-1} {\bf P}_b {\bf C}^{-1} {\bf P}_{b'}
  \right).
\label{eq:fish}
\end{equation}
The expectation and variance of the estimator are
\begin{equation}
\langle \hat{\bf \lambda} \rangle = {\bf A}{\bf F} {\bf \lambda}\,,
\end{equation}
\begin{equation}
\label{eq:var}
{\rm Var}(\hat{\bf \lambda},\hat{\bf \lambda}) = {\bf A} {\bf F} {\bf A}^T\,,
\end{equation}
and the window functions are ${\bf W} = {\bf A}{\bf F}$.  The matrix
${\bf A}$ can be chosen in a variety of ways to optimize the
estimator \cite{Tegmark97}.  Because the Fisher matrix is
ill-conditioned due to the small survey size, we use the robust
normalization $A_{ii} = (\sum_j F_{ij})^{-1}$.

The covariance between the $\Cl$ measurements of two redshift slices
labeled $A$ and $B$ may be estimated as
\begin{equation}
{\rm Var}(\Cl^A,\Cl^B)=\frac{1}{f_{sky}}\frac{2}{2\ell+1} \left(\Cl^{AB}\right)^2,
\end{equation}
where $\Cl^{AB}$ is given by eq. (\ref{eq:cl}), rewriting as
\begin{equation}
C_\ell^{AB} = \frac{2b_Ab_B}{\pi}\int k^2dkP(k)g^A_\ell(k)g^B_\ell(k)~.
\end{equation}
We must further convolve by the survey window functions.  This is an
idealization that neglects the precise survey geometry but we may
scale it to match the variance computed from the quadratic estimator

\begin{table}[t]
\caption{CFHTLS samples}\label{table:samples}
\begin{tabular}{c|c|c|c c}
\hline \hline
 \multicolumn{3}{c|}{}   &   \multicolumn{2}{c}{Scale ($\hmpc$)}  \\
 \multicolumn{2}{c}{Photo-z sample}  & ${\rm {\bar z}}$ & $\ell=630$ & $\ell=960$ \\
\hline
S6 & $0.5<{\rm z_{phot}} <0.6$ & 0.557 & 0.43 & 0.65 \\
S7 & $0.6<{\rm z_{phot}} <0.8$ & 0.687 & 0.36 & 0.55 \\
S8 & $0.8<{\rm z_{phot}} <1.0$ & 0.839 & 0.31 & 0.47 \\
\hline \hline
\end{tabular}
\end{table}

\subsection{$\Cl$ measurements}

The quadratic $\Cl$ estimator operates under the assumption that the
galaxy density can be described by a Gaussian random field, which is
not valid on small scales.  However, Gaussianity can be a reasonable
model for galaxy counts in projection even to non-linear scales. Our
map resolution is 7 arcmin/pixel ($\nside=512$) and, in the first
redshift bin, S6, the mean and median counts are 33.6 and 32.0
galaxies/pixel, respectively, with skewness 1.18.  At a higher
resolution of 3.5 arcmin/pixel, the mean and median become 8.42 and
7.72 with skewness 1.48.  We see that the map resolution of 7
arcmin/pixel gives a good balance between Gaussianity and angular
scale.  We expect that the Gaussianity assumption is still valid at
scales typical of the pixel size at $\ell=1500$.

We compute the angular power spectrum in 48 bins over the range
$\ell=2-960$, with width $\Delta\ell=20$. The Fisher matrix and
window functions were computed with step size $\Delta\ell=10$.  Due
to the small angular extent of the survey, we cannot probe low
$\ell$ and, indeed, the computed window functions show that we are
not sensitive to $\ell<40$.

\begin{figure}[htbp]
\begin{center}
\includegraphics[scale=0.3]{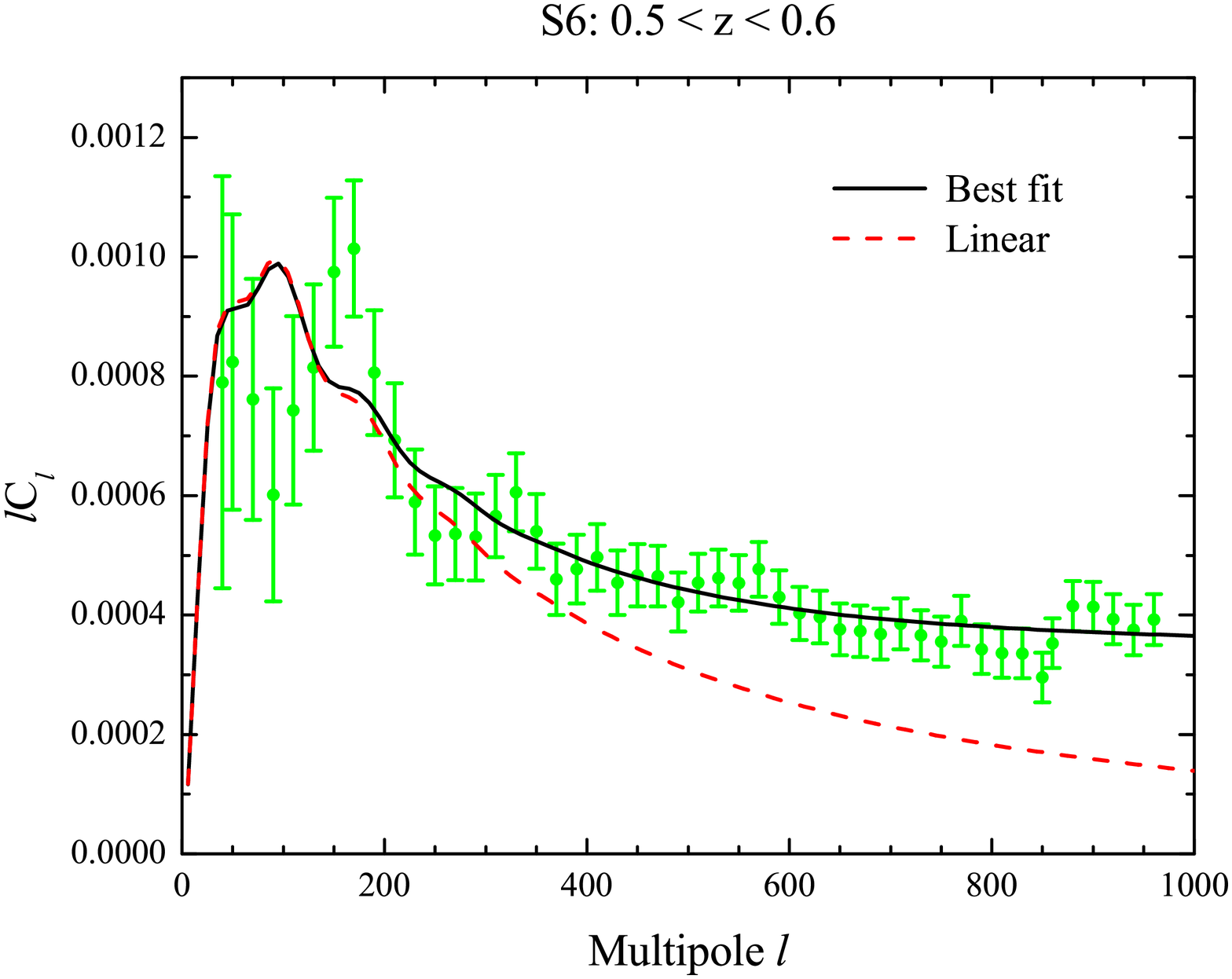}
\includegraphics[scale=0.3]{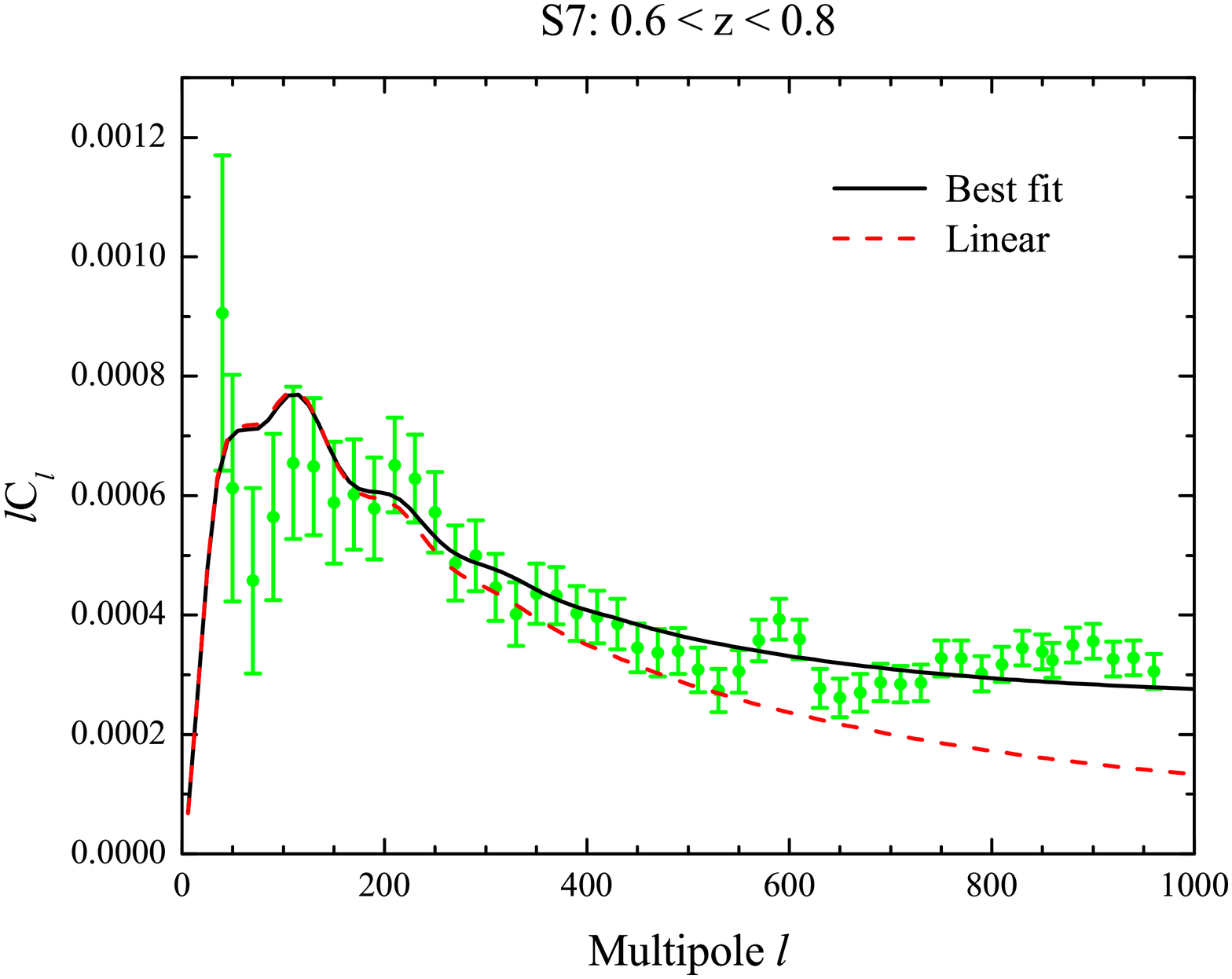}
\includegraphics[scale=0.3]{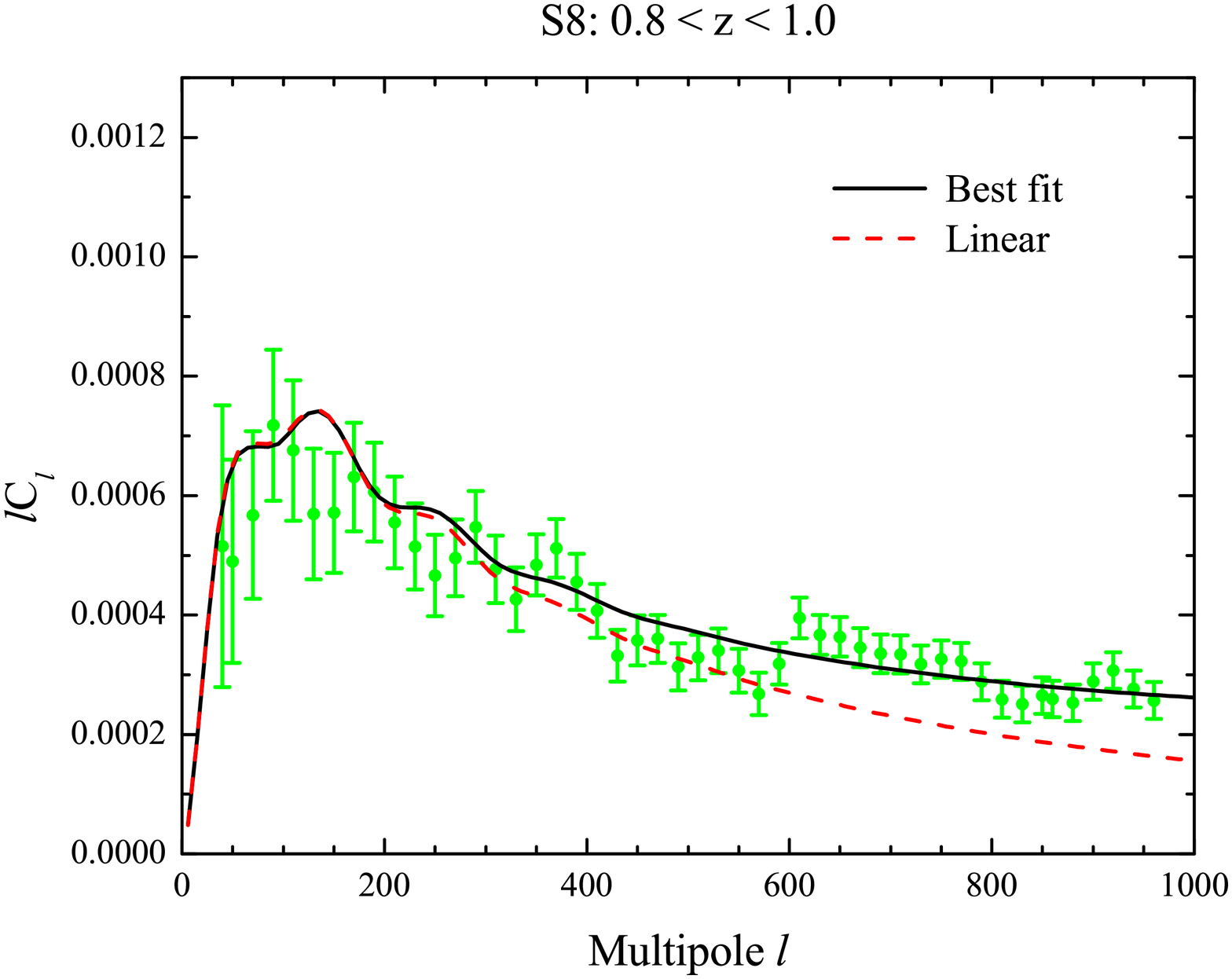}
\caption{The angular power spectra for the three different redshift
bins. We show the data with $1\,\sigma$ error bars and the linear
(dashed curves) and non-linear (continuous curves) theoretical
angular power spectra. \label{fig:cl}}
\end{center}
\end{figure}

\begin{figure}[htbp]
\begin{center}
\includegraphics[scale=1.5]{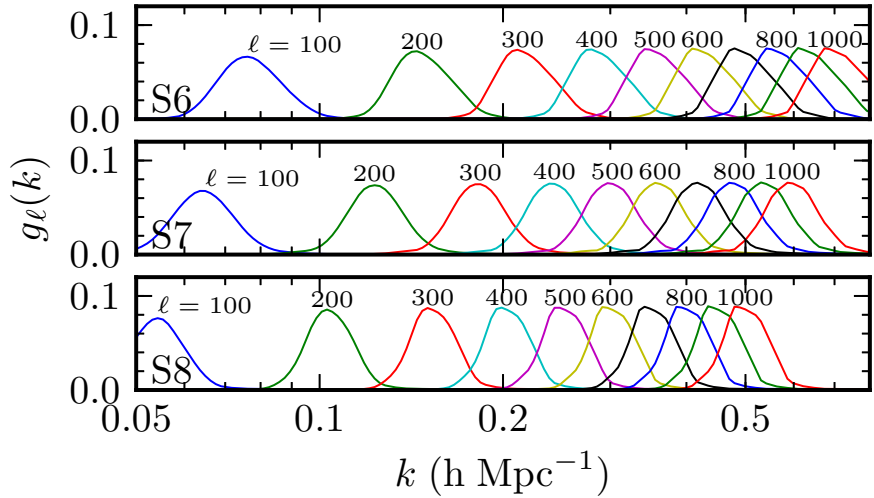}
\caption{The projection kernels $g_{\ell}(k)$ defined in eq.
(\ref{eq:gl}) for the three CFHTLS samples S6, S7 and S8 with mean
redshifts 0.557, 0.687 and 0.839.  The normalized kernels are
plotted. \label{fig:kernels}}
\end{center}
\end{figure}

We estimate $\Cl$ for the three photometric samples, S6, S7 and S8.
The measurements are shown in figure \ref{fig:cl}.  The Fourier
scales that are probed in the angular power spectrum depend on the
projection kernel $g_{\ell}$.  In figure \ref{fig:kernels}, we plot
representative kernels. We use two limits in angular scale in the
subsequent cosmological analysis: $\ell_{\rm max}=630$ and
$\ell_{\rm max}=960$.

\section{External Data Sets}\label{otherdata}

Besides the angular power spectra $C_\ell$ of the CFHTLS
measurement, we will also consider the following cosmological
probes: i) power spectra of CMB temperature and polarization
anisotropies; ii) power spectra of luminous red galaxies; iii)
measurement of the current Hubble constant; iv) luminosity distances
of type Ia supernovae. These data sets are all well established
cosmological probes that have been extensively investigated and
already provide tight constraints on the cosmological parameters of
the concordance $\Lambda$CDM model we will use here.

\subsection{CMB Power Spectra Data}

To incorporate the seven-year WMAP (WMAP7) CMB temperature and
polarization power spectra, we use the routines for computing the
likelihood supplied by the WMAP team \cite{Komatsu2011}. The WMAP7
polarization data are composed of TE/EE/BB power spectra on large
scales ($2\leq \ell\leq23$) and TE power spectra on small scales
($24\leq \ell \leq800$), while the WMAP7 temperature data includes
the CMB anisotropies on scales $2\leq\ell\leq1200$.

Here we do not use other small-scale CMB temperature power spectra
measurements, as adding them would not significantly improve the
constraints on the cosmological parameters, especially those on
the total neutrino mass focussed on in this paper.

\subsection{Power Spectrum of Luminous Red Galaxies}

The power spectrum of LRGs measured by SDSS is a powerful probe of
the total mass of neutrinos, $\sum{m_{\nu}}$, and the effective
number of neutrino species, $N_{\rm eff}$. We thus include the SDSS
DR4 LRG power spectrum \cite{Tegmark2006} which has a mean redshift
$\bar{z}\sim0.35$ and use data points on scales $0.012\,h/{\rm
Mpc}<k_{\rm eff}<0.203\,h/{\rm Mpc}$ in the analysis.

We checked the constraining power of the SDSS DR7 LRG power spectrum
\cite{Reid2010a} and found that the constraint on the total neutrino
mass does not improve significantly. For simplicity, we still use
the SDSS DR4 LRG power spectrum \cite{Tegmark2006}. Furthermore, we
do not include the BAO information \cite{Reid2010b}, as the
measurement of BAO and LRGs power spectrum cannot be treated as
independent data sets.

\subsection{Hubble Constant}

In our analysis, we add a Gaussian prior on the current Hubble
constant given by ref. \cite{Riess2009}; $H_0=74.2\pm3.6\,{\rm
km\,s^{-1}\,Mpc^{-1}}$ (68\% C.L.). The quoted error includes both
statistical and systematic errors. This measurement of $H_0$ is
obtained from the magnitude-redshift relation of 240 low-$z$ Type Ia
supernovae at $z<0.1$ by the Near Infrared Camera and Multi-Object
Spectrometer (NICMOS) Camera 2 of the Hubble Space Telescope (HST).
This is a significant improvement over the previous prior,
$H_0=72\pm8\,{\rm km\,s^{-1}\,Mpc^{-1}}$, which is from the Hubble
Key project final result \cite{Freedman2001}. In addition, we impose
a weak top-hat prior on the Hubble parameter: $H_0\in[40,100]\,{\rm
km\,s^{-1}\,Mpc^{-1}}$.

\subsection{Luminosity Distances}

Finally, we include data from Type Ia supernovae, which consists of
luminosity distance measurements as a function of redshift; $D_{\rm
L}(z)$. In this paper we use the latest SNIa data sets from the
Supernova Cosmology Project, ``Union Compilation 2.1'', which
consists of 580 samples and spans the redshift range $0\lesssim z
\lesssim 1.55$ \cite{Union2.1}. This data set also provides the
covariance matrix of data with and without systematic errors. In
order to be conservative, we use the covariance matrix with
systematic errors. When calculating the likelihood from SNIa, we
marginalize over the absolute magnitude $M$, which is a nuisance
parameter, as done in refs. \cite{Goliath2001,DiPietro2002}.

\begin{figure*}[t]
\begin{center}
\includegraphics[scale=0.5]{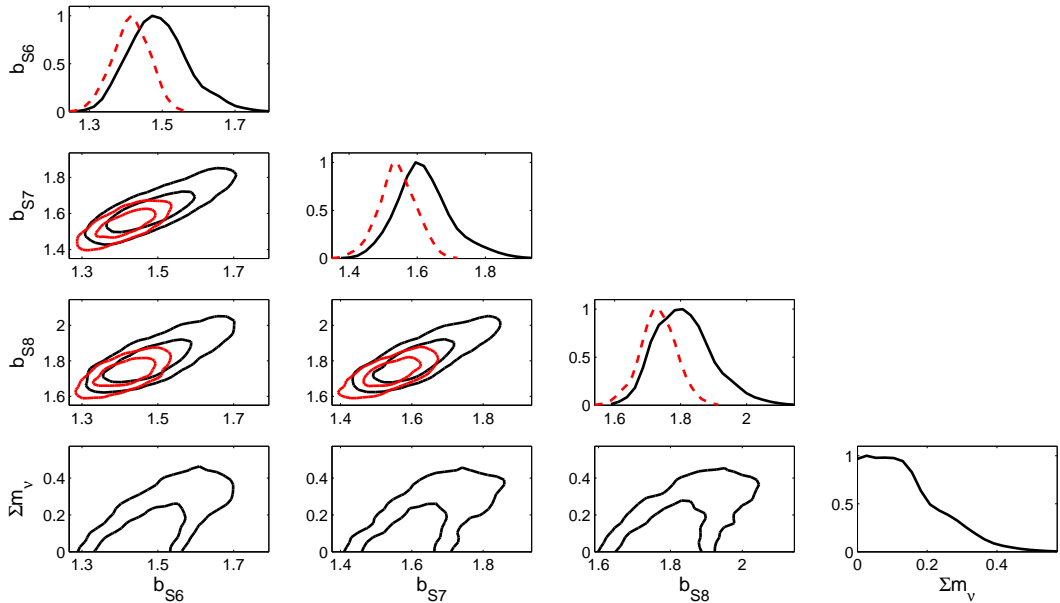}
\caption{Marginalized one-dimensional and two-dimensional likelihood
($1,\,2\,\sigma$ contours) constraints on the total neutrino mass
and the three CFHTLS bias parameters from WMAP7+HST+SDSS+SN+CFHTLS
data combination for $\ell_{\rm max}=630$. We also show the
constraints on the bias parameters, assuming massless neutrinos (red
dashed lines).\label{fig:bias}}
\end{center}
\end{figure*}

\section{Numerical Results}\label{results}

We model the nonlinear matter power spectrum $P(k)$ using the {\tt
Halofit} formulae \cite{Smith2003} as modified to account for
massive neutrinos \cite{Bird2012}. This modified version of {\tt
Halofit} was obtained from an extensive suite of N-body simulations
which treat massive neutrinos as an independent set of particles
\cite{vhs010}, and was particularly focused on exploring small
scales at redshifts $z=0-2$. Similar simulations have been used to
estimate the redshift space distortions and bias between matter and
haloes \cite{marulli}.

We perform a global fitting of cosmological parameters using the
{\tt CosmoMC} package \cite{cosmomc}, a Markov Chain Monte Carlo
(MCMC) code. We assume purely adiabatic initial conditions and a
flat $\Lambda$CDM Universe, with no tensor contribution to
primordial fluctuations. The following six cosmological parameters
are allowed to vary with top-hat priors: the dark matter energy
density parameter $\Omega_{\rm c} h^2 \in [0.01,0.99]$, the baryon
energy density parameter $\Omega_{\rm b} h^2 \in [0.005,0.1]$, the
primordial spectral index $n_{\rm s} \in [0.5,1.5]$, the primordial
amplitude $\log[10^{10} A_{\rm s}] \in [2.7,4.0]$, the ratio
(multiplied by 100) of the sound horizon at decoupling to the
angular diameter distance to the last scattering surface
$\Theta_{\rm s} \in [0.5,10]$, and the optical depth to reionization
$\tau \in [0.01,0.8]$. The pivot scale is set at $k_{\rm
s0}=0.05\,$Mpc$^{-1}$. Besides these six basic cosmological
parameters, we pay particularly attention to the neutrino mass
fraction at the present day
\begin{equation}
f_\nu\equiv\frac{\Omega_\nu{h^2}}{\Omega_m{h^2}}=\frac{\sum{m_{\nu}}}{93.14\,{\rm eV}\,\Omega_mh^2}~,
\end{equation}
the effective number of neutrino species, $N_{\rm eff}$, and the
three CFHTLS galaxy bias parameters, $b_{\rm S6}$, $b_{\rm S7}$,
$b_{\rm S8}$. Note that in our analyses we do not vary $\sum{m_\nu}$
and $N_{\rm eff}$ simultaneously, since they are no longer
degenerate \cite{Hannestad2006}. Instead, we assume $N_{\rm
eff}=3.04$ to constrain the total mass of neutrinos and
$\sum{m_\nu}=0$ to constrain $N_{\rm eff}$.

\begin{table*}[t]
\caption{The 95\% confidence level upper limits on the total mass of
neutrinos $\sum{m_\nu}$ from different data
combinations.}\label{table:mnu}
\begin{tabular}{l|cc|cc}
\hline \hline
95\% C.L. $\sum{m_\nu}$ [eV]  & \multicolumn{2}{c|}{Without HST Prior} & \multicolumn{2}{c}{With HST Prior} \\
\cline{2-5}
&$\ell_{\rm max}=630$ & $\ell_{\rm max}=960$&$\ell_{\rm max}=630$ & $\ell_{\rm max}=960$\\
\hline

WMAP7 & \multicolumn{2}{c|}{1.17} & \multicolumn{2}{c}{0.50}\\
WMAP7 + CFHTLS & 0.64 &0.43& 0.41& 0.29\\
WMAP7 + SDSS + CFHTLS & 0.47 &0.35& 0.35& 0.28\\
WMAP7 + SDSS + SN + CFHTLS & $-$ & $-$& 0.33 &0.27\\

\hline  \hline
\end{tabular}
\end{table*}

\subsection{CFHTLS galaxy bias}

We first measure the galaxy bias in the context of the $\Lambda$CDM
model, without introducing massive neutrinos.  We find the 68\% C.L.
constraints on the bias parameters to be: $b_{\rm S6}=1.41\pm 0.05$,
$b_{\rm S7}=1.54\pm 0.05$ and $b_{\rm S8}=1.73\pm 0.06$, which are
in good agreement with those derived by ref.  \cite{Granett2012}.
The bias values are consistent, but higher than those found by ref.
\cite{Coupon012} for volume limited samples in CFHTLS covering the
same redshift ranges.  This discrepancy could arise from differences
in the analysis method; in ref. \cite{Coupon012}, the cosmology was
fixed and the fit was carried out over halo model parameters using
the correlation function on small scales ($<1^{\circ}$).  Generally,
the best-fitting bias values are correlated with the other
parameters considered in the analysis.  In particular, we will find
that they shift when we introduce massive neutrinos.

We also check whether a scale-dependent bias is supported by the
data, again assuming $\Lambda$CDM without massive neutrinos. We
compute the best-fitting parameters assuming a scale-dependent $Q$
bias model \cite{Cole2005} with the form
\begin{equation}
b(k) = b_{\rm lin}\sqrt{\frac{1+Qk^2}{1+Ak}}.
\end{equation}
We fix the parameter $A=1.7\,$Mpc/h and find the best-fitting $Q$.

We first perform the fit using the linear power spectrum, such that
$P(k) = b^2(k)P_{\rm lin}(k)$.  The best-fitting $Q$ value is
$Q_{\rm lin}=10.3\pm1.5$.  If we instead use a non-linear {\tt
Halofit} power spectrum (without massive neutrinos), the
best-fitting $Q$ value is reduced to $Q_{\rm halofit}=2.5\pm1.2$ and
scale dependence in the bias at the $<5\%$ level at $k<0.6\hmpc$,
which is negligible. The robust fit using {\tt Halofit} and a simple
constant bias is expected, since the data is well-fit by the
two-halo clustering term on these scales \cite{Coupon012}. This
demonstrates that the assumption of constant galaxy bias is a good
one, and our results for massive neutrinos are robust.

Before presenting the constraints on the total neutrino mass, we
examine the degeneracies between the total neutrino mass and the
three CFHTLS bias parameters. In figure \ref{fig:bias} we show the
marginalized one-dimensional and two-dimensional likelihood
constraints from the WMAP7+HST+SDSS+SN+CFHTLS data combination. The
black solid lines and the red dashed lines denote the constraints
with massless and massive neutrinos, respectively. When including
massive neutrinos, the constraints on the bias parameters are
obviously weakened and the median values are shifted: $b_{\rm
S6}=1.49 \pm 0.07$, $b_{\rm S7}=1.62 \pm 0.08$ and $b_{\rm S8}=1.81
\pm0.08$ (68\% C.L.). This results in a strong correlation between
$\sum{m_\nu}$ and the bias parameters; clearly shown in the
two-dimensional contours of the last row of figure \ref{fig:bias}. A
larger total neutrino mass will suppress the amplitude of the matter
power spectrum on small scales further. Countering this while still
matching the CFHTLS data on small scales requires larger bias
parameters. Thus, further improving the constraints on the total
neutrino mass will require better determination of the bias
parameters.

\begin{figure*}[htbp]
\begin{center}
\includegraphics[scale=0.5]{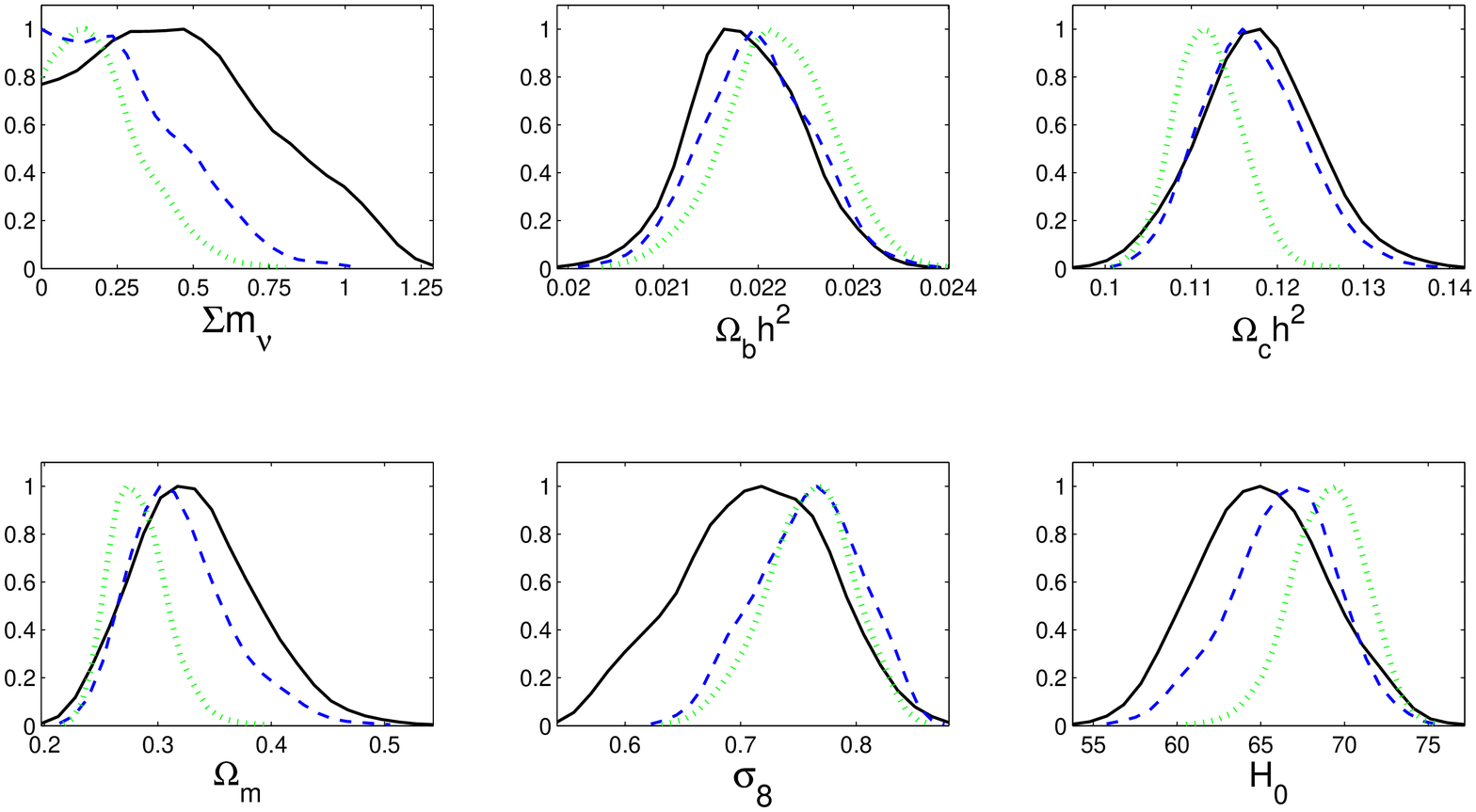}
\caption{One-dimensional marginalized likelihood constraints on the
total neutrino mass $\sum{m_\nu}$, as well as other cosmological
parameters, $\Omega_bh^2$, $\Omega_ch^2$, $\Omega_m$, $\sigma_8$ and
$H_0$ from different data combinations: WMAP7 alone (black solid
lines), WMAP7+CFHTLS (blue dashed lines) and WMAP7+SDSS+CFHTLS
(green dotted lines) for $\ell_{\rm max}=630$.\label{fig:WMAP7-1d}}
\end{center}
\end{figure*}

\begin{figure*}[t]
\begin{center}
\includegraphics[scale=0.4]{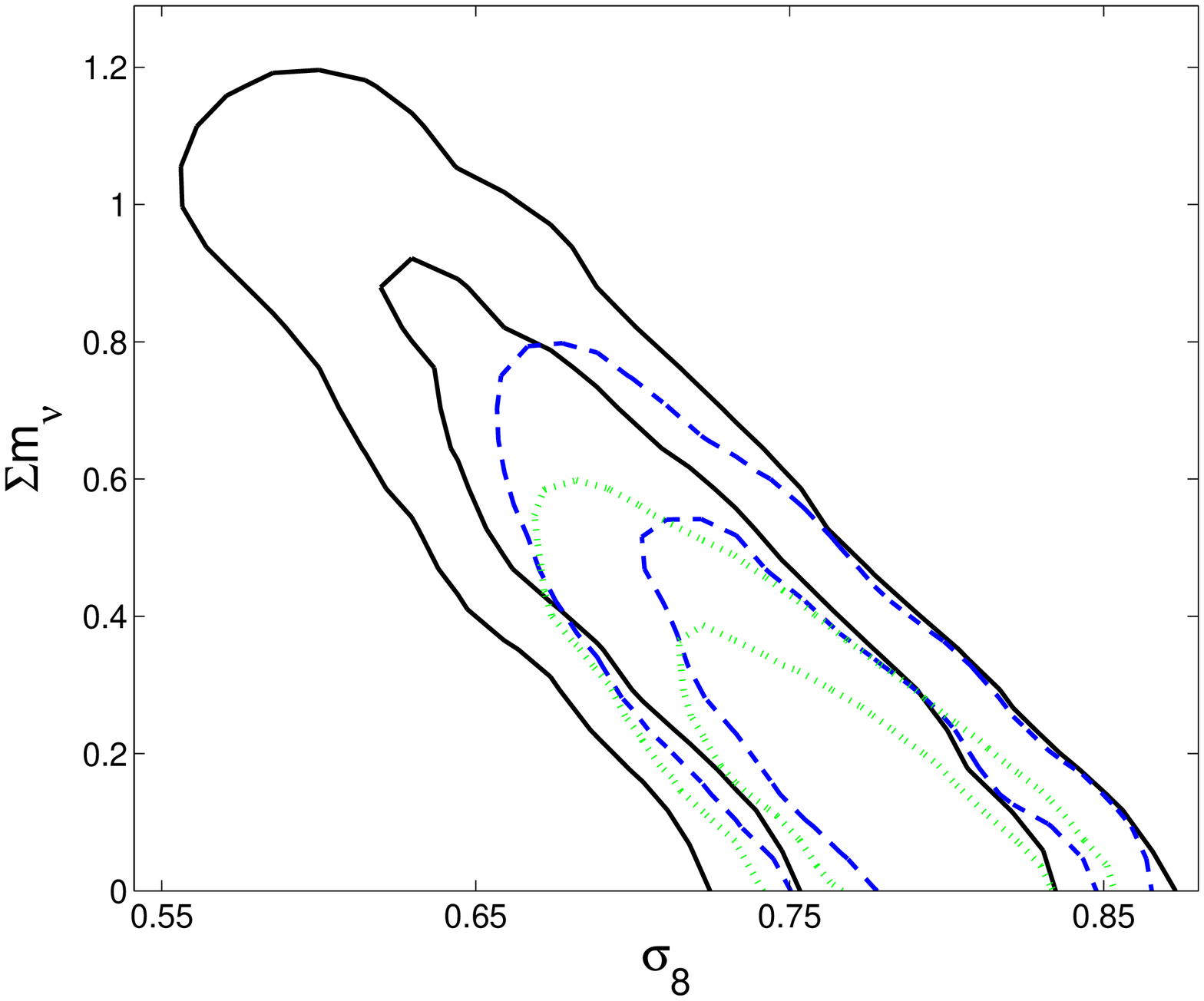}
\includegraphics[scale=0.4]{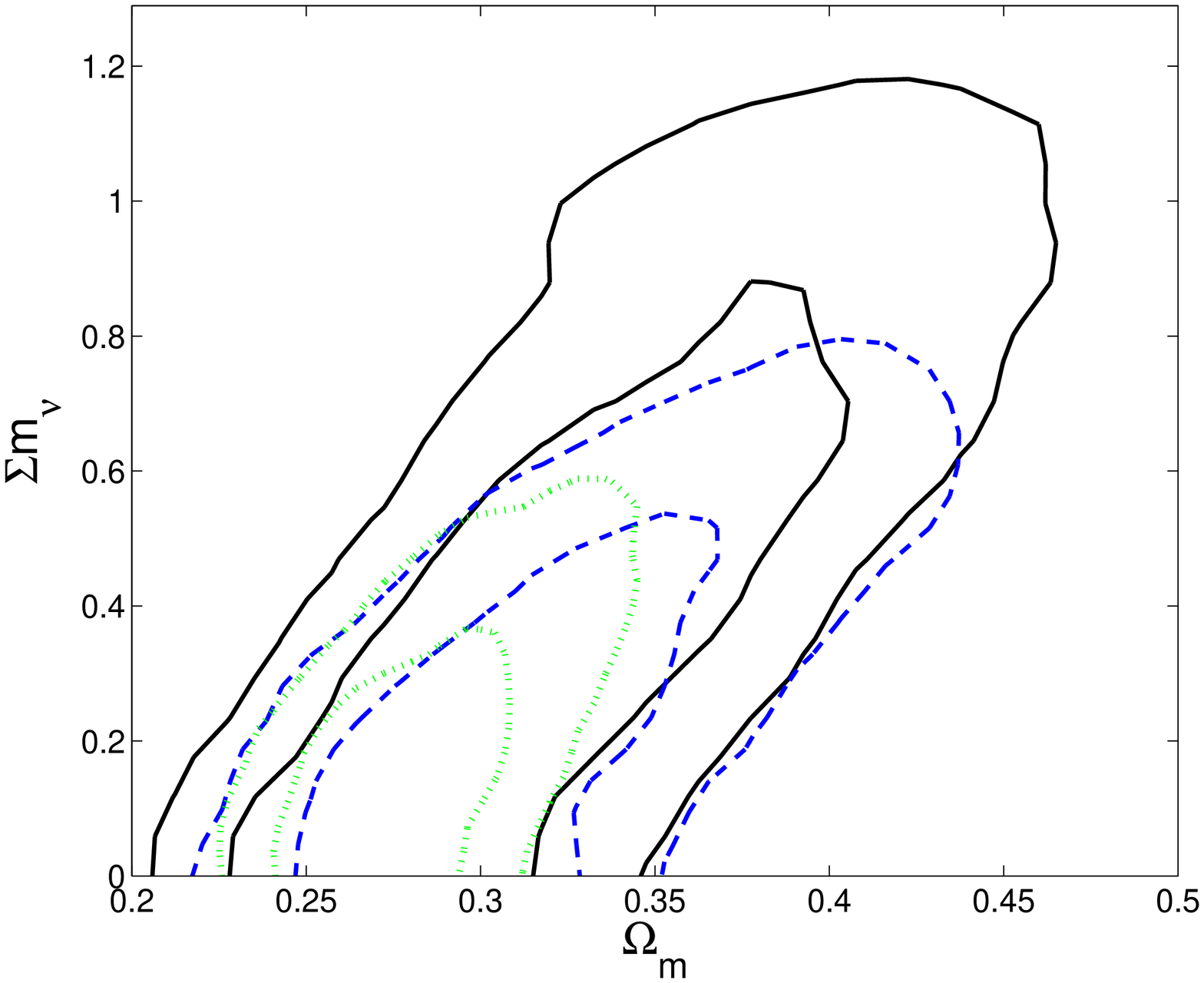}
\includegraphics[scale=0.4]{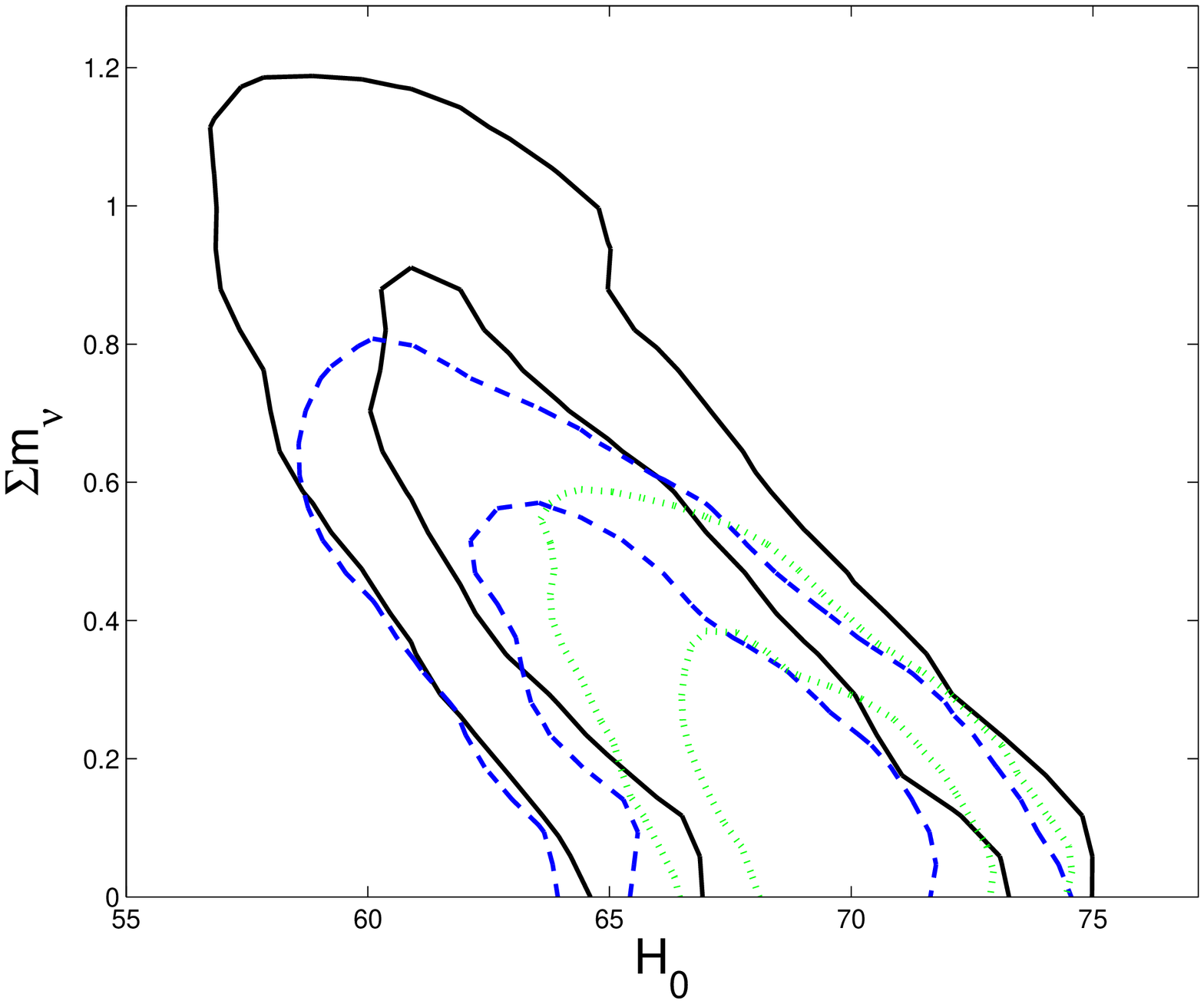}
\caption{Two-dimensional contours in the ($\sigma_8$,$\sum{m_\nu}$),
($\Omega_m$,$\sum{m_\nu}$) and ($H_0$,$\sum{m_\nu}$) panels from
different data combinations: WMAP7 alone (black solid lines),
WMAP7+CFHTLS (blue dashed lines) and WMAP7+SDSS+CFHTLS (green dotted
lines) for $\ell_{\rm max}=630$.\label{fig:WMAP7-2d}}
\end{center}
\end{figure*}

\subsection{Neutrino Mass $\sum{m_{\nu}}$}

In this subsection, we present the 95\% confidence level upper
limits on the total mass of neutrinos from different data
combinations after marginalizing over the other parameters, as shown
in table \ref{table:mnu}.

We start by presenting the limits using scales up to $\ell_{\rm
max}=630$. Due to the strong degeneracies present between
cosmological parameters, primary CMB anisotropies alone can place
only weak constraints on the total neutrino mass. In the flat
$\Lambda$CDM framework, WMAP7 alone constrains
$\sum{m_\nu}<1.17\,$eV at the 95\% confidence level
\cite{Komatsu2011}. Adding the low redshift CFHTLS measurements
breaks these degeneracies and the constraint on the total neutrino
mass significantly improves to
\begin{equation}
\sum{m_\nu}<0.64\,{\rm eV}~~~(95\%~{\rm C.L.})
\end{equation}
for the combined WMAP7 and CFHTLS datasets. In figure
\ref{fig:WMAP7-1d}, we plot the one-dimensional marginalized
distributions on some cosmological parameters from different data
combinations. As can be seen, adding CFHTLS data improves the
constraints on the present matter density $\Omega_m$, the amplitude
of fluctuations $\sigma_8$ and the hubble parameter $H_0$; their
68\% confidence levels are shrunk from $0.334\pm0.053$,
$0.711\pm0.062$ and $65.1\pm3.7$ for WMAP7 alone to $0.320\pm0.043$,
$0.759\pm0.044$ and $66.3\pm3.1$ for WMAP7+CFHTLS data,
respectively. In figure \ref{fig:WMAP7-2d} we show the
two-dimensional contours in the ($\sigma_8$,$\sum{m_\nu}$),
($\Omega_m$,$\sum{m_\nu}$) and ($H_0$,$\sum{m_\nu}$) planes from
different data combinations. Using WMAP7+CFHTLS data (blue dashed
lines) reduces the correlations between $\sum{m_\nu}$ and other
parameters and gives tighter constraints on $\sum{m_\nu}$ than WMAP7
alone (black solid lines).

Using the SDSS DR4 LRG power spectrum data \cite{Tegmark2006}, we
obtain a constraint on the total neutrino mass of
$\sum{m_\nu}<0.62\,$eV at the 95\% confidence level, comparable to
the limit from WMAP7+CFHTLS. For comparison, we also use the SDSS
DR7 LRG power spectra data \cite{Reid2010a} to constrain
$\sum{m_\nu}$ and get a 95\% C.L. upper limit of
$\sum{m_\nu}<0.61\,$eV, which is almost identical with that from DR4
LRG data. Due to this negligible improvement, in our calculations we
still use the SDSS DR4 data to avoid using the more complicated DR7
likelihood code. Figure \ref{fig:WMAP7-1d} (green dashed lines)
clearly shows that the constraints on $\Omega_m$, $\sigma_8$ and
$H_0$ are much tighter when including the SDSS data set. We obtain a
constraint on the total neutrino mass of $\sum{m_\nu}<0.47\,$eV at
the 95\% confidence level from the combination of WMAP7+SDSS+CFHTLS
data.


\begin{figure*}[t]
\begin{center}
\includegraphics[scale=0.5]{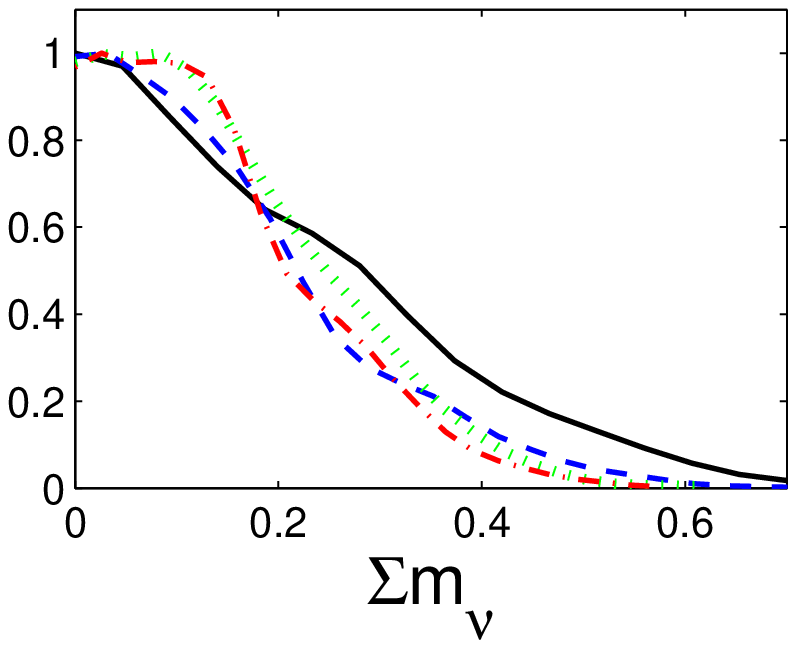}
\includegraphics[scale=0.5]{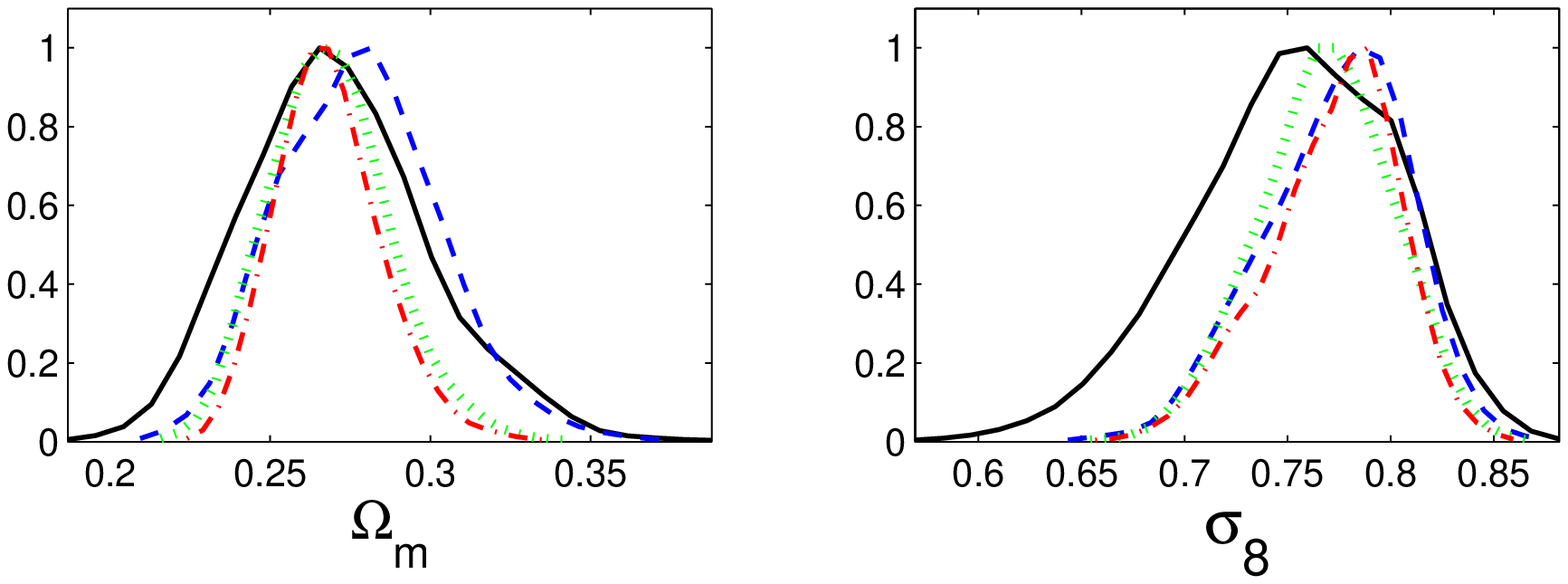}
\caption{One-dimensional marginalized likelihood constraints on the
total neutrino mass $\sum{m_\nu}$, as well as other cosmological
parameters, $\Omega_m$ and $\sigma_8$ from different data
combinations: WMAP7+HST (black solid lines), WMAP7+HST+CFHTLS (blue
dashed lines), WMAP7+HST+SDSS+CFHTLS (green dotted lines) and
WMAP7+HST+SDSS+SN+CFHTLS (red dash-dot lines) for $\ell_{\rm
max}=630$.\label{fig:WMAP7HST-1d}}
\end{center}
\end{figure*}

We next consider the constraints using WMAP7 with an HST prior. The
degeneracy between $H_0$ and $\sum{m_\nu}$, clearly shown in figure
\ref{fig:WMAP7-2d}, means that independent measurements of  $H_0$
produce significantly tighter overall constraints. WMAP7+HST gives
the 95\% C.L. upper limit $\sum{m_\nu}<0.50\,$eV. After adding the
low redshift CFHTLS clustering power spectra data, the constraint
becomes even tighter
\begin{equation}
\sum{m_\nu}<0.41\,{\rm eV}~~~(95\%~{\rm C.L.})~.
\end{equation}
In figure \ref{fig:WMAP7HST-1d} we show the one-dimensional
marginalized distributions on some cosmological parameters from
different data combinations when the HST prior is used. Clearly, the
CFHTLS data improves the constraints on some cosmological
parameters, reducing the correlations between them and $\sum{m_\nu}$
further, and thus significantly improving the limit on
$\sum{m_\nu}$.

For comparison, we use the SDSS DR4 LRG power spectra data to
constrain $\sum{m_\nu}$ and obtain the 95\% upper limit $0.45\,$eV,
comparable to that from the CFHTLS data set. Similarly, we use the
SDSS DR7 LRG data and obtain the constraint $\sum{m_\nu}<0.43\,$eV
(95\% C.L.), which is consistent with previous work
\cite{Komatsu2011}. When using SDSS and CFHTLS data sets together,
we get tighter constraints on the total neutrino mass;
$\sum{m_\nu}<0.35\,$eV at the 95\% confidence level. Finally, we add
the ``Union 2.1 Compilation'' supernovae data into the
WMAP7+HST+SDSS+CFHTLS data combination and the constraint on the
total neutrino mass becomes
\begin{equation}
\sum{m_\nu}<0.33\,{\rm eV}~~~(95\%~{\rm C.L.})~.
\end{equation}

We now present the constraints obtained from extending the ranges of
multipoles to smaller scales; up to $\ell_{\rm max}=960$. The
improvements from these scales are not negligible: if we consider
CFHTLS data in combination with WMAP7 we go from $\sum{m_\nu}=0.64$
eV to $\sum{m_\nu}<0.43$ eV. If we also add the HST data we obtain
$\sum{m_\nu}<0.29$ eV, which has to be compared with the $\ell_{\rm
max}=630$ results of $\sum{m_\nu}=0.41$ eV. Note that by adding SDSS
and SN data the constraint improves and becomes $\sum{m_\nu}<0.27$
eV at the $2\sigma$ C.L. level, which is the tightest bound on
neutrinos presented in this paper.

It is evident that exploring the smaller scale matter power spectrum
can significantly improve constraints obtained from larger scales.
Our result above can be compared to the constraint obtained using
only multipoles up to the linear theory value $\ell_{\rm max}=150$,
which is $\sum{m_\nu}<0.43$ eV for the combined data set.

Finally, we have explored the effect that the error on the
non-linear fitting formula for the matter power has on the final
results. This error has been quantified in ref. \cite{Bird2012},
which presents an analytic expression, $E(k,z)$, the error on the
suppression in the power spectrum due to neutrinos, which depends on
scale, redshift and neutrino mass fraction, and is at the level of
$5$\%. We computed the constraints by considering
$P(k)=P(k)[1+aE(k,z)]$, with $a$ being a free parameter with
standard deviation of unity. For all data sets the $2\,\sigma$ upper
limit for $\ell_{\rm max}=960$ weakens by only $0.01$eV,
demonstrating that the numerical errors for the scales considered
here are smaller than statistical uncertainties on the data point
and will not impact the final constraints. {\tt Halofit} will also
underestimate the matter power spectrum, in a manner independent of
neutrino mass, by around $5$\% \cite{Heitmann2010}. Neglecting this
error, because it underestimates the matter power spectrum, is
conservative and will only produce slightly weaker constraints.
Furthermore, since the magnitude of this error is similar to
$E(k,z)$ and independent of neutrino mass, its effect should be
small.

\begin{table*}[htbp]
\caption{The 68\% and 95\% confidence levels on the effective number
of neutrino species, $N_{\rm eff}$ from different data
combinations.}\label{table:neff}
\begin{tabular}{l|cc}
\hline \hline
Data sets & \multicolumn{2}{c}{$N_{\rm eff}$}\\
\cline{2-3}
&$\ell_{\rm max}=630$ & $\ell_{\rm max}=960$\\
\hline

WMAP7 alone & \multicolumn{2}{c}{$> 2.84~(95\%\,{\rm C.L.})$}\\
WMAP7 + SDSS & \multicolumn{2}{c}{$4.13^{+1.20}_{-0.89}(^{+4.05}_{-2.00})$}\\
WMAP7 + CFHTLS & $4.38 ^{+2.04}_{-1.16} (^{+3.72}_{-2.12})$&$4.23 ^{+1.55}_{-0.70} (^{+2.96}_{-1.50})$\\

\hline

WMAP7 + HST & \multicolumn{2}{c}{$3.99^{+1.76}_{-0.41}(^{+2.92}_{-1.36})$}\\
WMAP7 + HST + SDSS & \multicolumn{2}{c}{$3.93^{+0.85}_{-0.73}(^{+1.78}_{-1.42})$}\\
WMAP7 + HST + CFHTLS & $3.98^{+1.04}_{-0.51}(^{+2.02}_{-1.20})$&$4.17^{+0.80}_{-0.67}(^{+1.62}_{-1.26})$\\
WMAP7 + HST + SDSS + CFHTLS & $3.91^{+0.71}_{-0.68}(^{+1.38}_{-1.20})$& $3.92^{+0.75}_{-0.51}(^{+1.42}_{-1.03})$\\
WMAP7 + HST + SDSS + SN + CFHTLS & $3.92^{+0.61}_{-0.60}(^{+1.33}_{-1.17})$& $3.91^{+0.67}_{-0.45}(^{+1.26}_{-0.96})$\\

\hline  \hline
\end{tabular}
\end{table*}

\begin{figure*}[t]
\begin{center}
\includegraphics[scale=0.5]{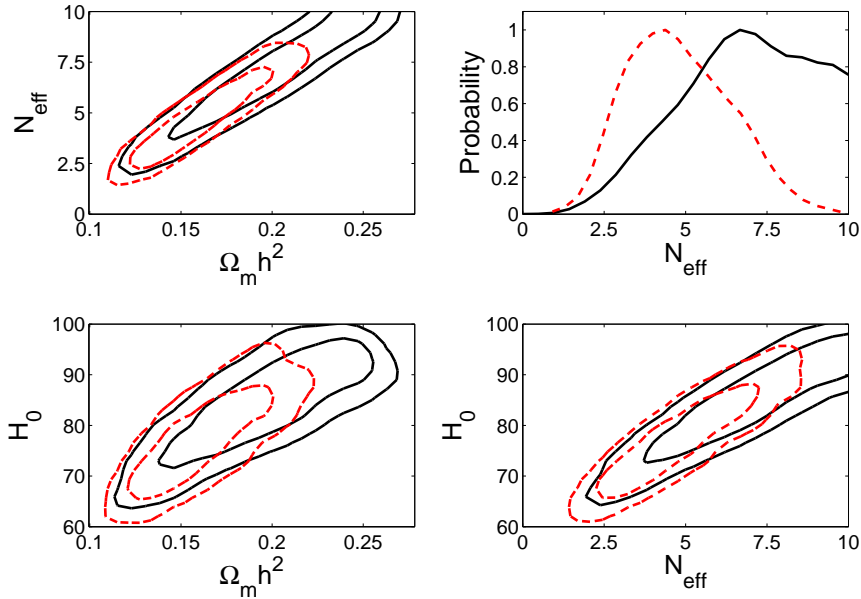}
\caption{Marginalized one-dimensional and two-dimensional likelihood
($1,\,2\,\sigma$ contours) constraints on $\Omega_mh^2$, $H_0$ and
$N_{\rm eff}$ from WMAP7 (black solid lines) and WMAP7+CFHTLS (red
dashed lines) data combinations for $\ell_{\rm max}=630$.
\label{fig:neff-2d}}
\end{center}
\end{figure*}

\subsection{Relativistic Species $N_{\rm eff}$}

In this subsection, we consider the constraints on the effective
number of neutrino species, $N_{\rm eff}$, assuming massless
neutrinos. Since $N_{\rm eff}$ can be written in terms of $\Omega_m
h^2$ and the redshift of matter-radiation equality, $z_{\rm eq}$,
there are strong degeneracies present between $N_{\rm eff}$, the
matter density, $\Omega_m h^2$ and the Hubble parameter $H_0$
\cite{Komatsu2009, Reid2010c}. CMB constraints on $N_{\rm eff}$ can
thus be strongly improved by combining them with measurements of the
small-scale matter power spectrum such as those obtained from SDSS
or CFHTLS.

We find the WMAP7 data alone gives $N_{\rm eff} > 2.84$ at the 95\%
confidence level, consistent with the result derived by the WMAP7
team \cite{Komatsu2011}. Adding the CFHTLS data significantly
improves the constraints on $N_{\rm eff}$ to
\begin{equation}
N_{\rm eff} = 4.38 ^{+2.04}_{-1.16} (^{+3.72}_{-2.12})
\end{equation}
at the 68\% and 95\% C.L.  Figure \ref{fig:neff-2d} shows the two
dimensional likelihood contours for both these constraints, with
$\ell_{\rm max} = 630$, making it clear that the improvement is
coming from degeneracy breaking. For comparison, when we use the
SDSS DR4 LRG power spectra data instead of CFHTLS, we obtain the
very consistent $N_{\rm eff}=4.13^{+1.20}_{-0.89}(^{+4.05}_{-2.00})$
(68 and 95\% C.L.).

\begin{figure*}[t]
\begin{center}
\includegraphics[scale=0.5]{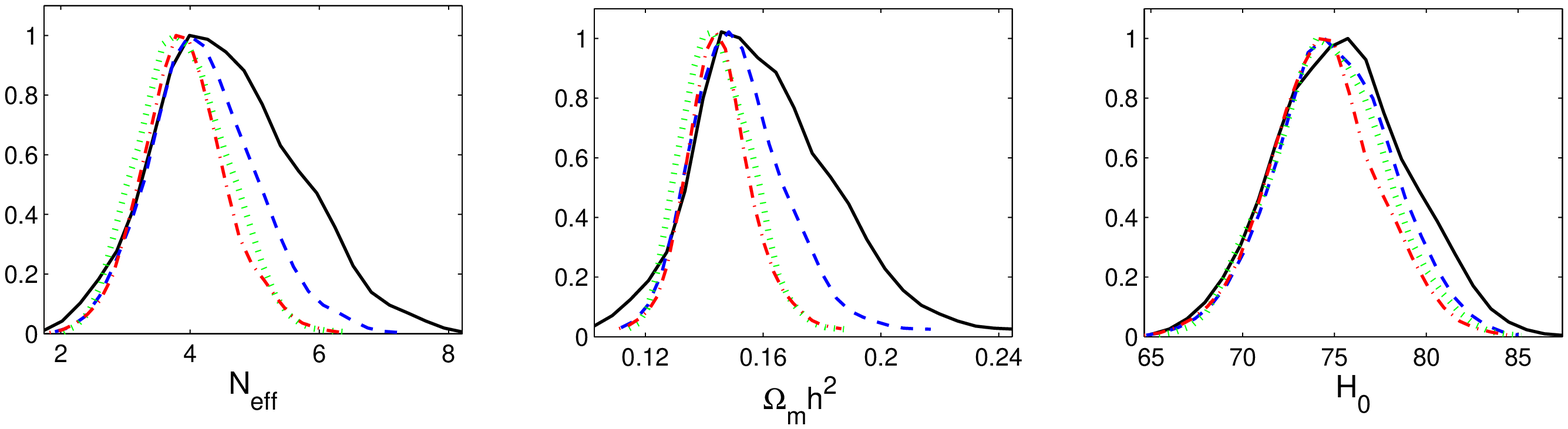}
\caption{One-dimensional marginalized likelihood constraints on
$N_{\rm eff}$, $\Omega_mh^2$ and $H_0$ from different data
combinations: WMAP7+HST (black solid lines), WMAP7+HST+CFHTLS (blue
dashed lines), WMAP7+HST+SDSS+CFHTLS (green dotted lines) and
WMAP7+HST+SDSS+SN+CFHTLS (red dash-dot lines) for $\ell_{\rm
max}=630$. \label{fig:neff-1d}}
\end{center}
\end{figure*}

WMAP7+HST gives $N_{\rm eff}=3.99^{+1.76}_{-0.41}(^{+2.92}_{-1.36})$
(68 and 95\% C.L.). Adding the CFHTLS angular power spectra tightens
this to $N_{\rm eff}=3.98^{+1.04}_{-0.51}(^{+2.02}_{-1.20})$. This
is similar to WMAP7+HST+SDSS, which gives $N_{\rm
eff}=3.93^{+0.85}_{-0.73}(^{+1.78}_{-1.42})$. WMAP7+HST+SDSS+CFHTLS
gives $N_{\rm eff}=3.91^{+0.71}_{-0.68}(^{+1.38}_{-1.20})$. Table
\ref{table:neff} lists these constraints on $N_{\rm eff}$, while
figure \ref{fig:neff-1d} shows one-dimensional marginalized
likelihood constraints on $N_{\rm eff}$, $\Omega_mh^2$ and $H_0$,
with $\ell_{\rm max}=630$.

WMAP7+HST+SDSS+CFHTLS+SN gives our most stringent constraint of
\begin{equation}
N_{\rm eff} = 3.92^{+0.61}_{-0.60}(^{+1.33}_{-1.17})~~~(68\%~{\rm
and}~95\%~{\rm C.L.})~.
\end{equation}
We also consider the $\ell_{\rm max}=960$ case and find that the
constraint improves slightly: $N_{\rm
eff}=3.91^{+0.67}_{-0.45}(^{+1.26}_{-0.96})$ for the 68\% and 95\%
confidence levels. Our results are quite consistent with those of
the ACT and SPT CMB experiments, which are $N_{\rm eff}=5.3\pm1.3$
and $N_{\rm eff}=3.85\pm0.62$ ($68$\% confidence level), and, like
them, display a slight preference for an extra relativistic relic
\cite{Dunkley2011,Keisler2011}. However, the standard value of
$N_{\rm eff}=3.04$ remains well within the 95\% confidence
intervals.

\section{Conclusions and Discussions}\label{summary}

Measurements of the galaxy power spectrum can put strong constraints
on the neutrino mass and effective number of neutrino species.
Significant gains can be made by probing weakly non-linear scales
and investigating redshift evolution. We have used the improved
parametrization developed by ref. \cite{Bird2012}, who
precisely calibrated the effect of massive neutrinos on the matter
power spectrum using a suite of N-body simulations, to investigate
non-linear scales.

Although the non-linear matter power spectrum may be estimated
accurately from simulations, connecting it to the galaxy clustering
is challenging.  In this study, we have assumed a simple bias model
with no dependence on scale.  To check the validity of this model,
we fit the power spectrum with a simple two-parameter $Q$ bias model
in $\Lambda$CDM without massive neutrinos.  We find that the
best-fitting $Q$ model has negligible scale-dependence to
$k\sim0.6\hmpc$.  We stress that this is only a consistency check,
but it demonstrates that a scale-dependent bias model is not needed
to fit the data at the redshifts we consider, $z>0.5$, even to
weakly non-linear scales of $k\sim0.6\hmpc$.  These results are in
part due to the range of galaxy luminosities represented in the
CFHTLS sample.  The CFHTLS data set thus probes an interesting
regime for neutrino studies for $z=0.5-1.2$ at small scales that can
be modeled with confidence.

We find that combining the CFHTLS data with WMAP already gives tight
limits on the total neutrino mass.  Adding additional low-redshift
probes, including SDSS LRGs and SN, only marginally improves the
constraints.  We benefit considerably by extending the angular scale
limit from $\ell_{\rm max}=630$ to $960$. This extends the fit from
$k\sim0.4\hmpc$ to $k\sim0.6\hmpc$.  However, there is a strong
correlation between galaxy bias and total neutrino mass, since both
parameters modulate the amplitude of the galaxy power spectrum.
Thus, future analyses can benefit from adding additional observables
to constrain the galaxy bias.

We further constrain the effective number of neutrino species.
Again, CFHTLS complements the WMAP7 data by constraining $\Omega_m$
leading to tighter limits on $N_{\rm eff}$.  We find that the
constraining power of the CFHTLS data is similar to that of the SDSS
LRG sample and the limits from both surveys are fully consistent.

The main constraints derived in this paper are summarized as
follows: CFHTLS, combined with WMAP7 and a prior on the Hubble
constant provides an upper limit of $\sum{m_\nu}<0.29\,$eV and
$N_{\rm{eff}} =4.17^{+1.62}_{-1.26}$ (2$\,\sigma$ confidence
levels). If we  instead omit smaller scales which may be affected by
non-linearities, these constraints relax to the following values:
$\sum{m_\nu}<0.41\,$eV and $N_{\rm{eff}} =3.98^{+2.02}_{-1.20}$
(2$\,\sigma$ confidence levels). By combining with the SDSS LRG matter
power and SN luminosity distance moduli we further improve to
$\sum{m_\nu}<0.27\,$eV and $N_{\rm{eff}} =3.91^{+1.26}_{-0.96}$
(2$\,\sigma$ confidence levels).

\section*{Acknowledgments}
CFHTLS is based on observations obtained with MegaPrime/MegaCam, a
joint project of CFHT and CEA/DAPNIA, at the Canada-France-Hawaii
Telescope (CFHT) which is operated by the National Research Council
(NRC) of Canada, the Institut National des Sciences de l'Univers of
the Centre National de la Recherche Scientifique (CNRS) of France,
and the University of Hawaii. This work is based in part on data
products produced at TERAPIX and the Canadian Astronomy Data Centre
as part of the Canada-France-Hawaii Telescope Legacy Survey, a
collaborative project of NRC and CNRS.

We acknowledge the use of the Legacy Archive for Microwave
Background Data Analysis (LAMBDA). Support for LAMBDA is provided by
the NASA Office of Space Science. MV is supported by ASI/AAE,
INFN/PD-51, PRIN MIUR, PRIN INAF 2009 and the European Research
Council - Starting Grant ``cosmoIGM''. SPB is supported by NSF grant
AST-0907969 and the Institute for Advanced Study. Funding for the
SDSS and SDSS-II has been provided by the Alfred P. Sloan
Foundation, the Participating Institutions, the National Science
Foundation, the U.S. Department of Energy, the National Aeronautics
and Space Administration, the Japanese Monbukagakusho, the Max
Planck Society, and the Higher Education Funding Council for
England. The SDSS Web Site is http://www.sdss.org/. The SDSS is
managed by the Astrophysical Research Consortium for the
Participating Institutions. The Participating Institutions are the
American Museum of Natural History, Astrophysical Institute Potsdam,
University of Basel, University of Cambridge, Case Western Reserve
University, University of Chicago, Drexel University, Fermilab, the
Institute for Advanced Study, the Japan Participation Group, Johns
Hopkins University, the Joint Institute for Nuclear Astrophysics,
the Kavli Institute for Particle Astrophysics and Cosmology, the
Korean Scientist Group, the Chinese Academy of Sciences (LAMOST),
Los Alamos National Laboratory, the Max-Planck-Institute for
Astronomy (MPIA), the Max-Planck-Institute for Astrophysics (MPA),
New Mexico State University, Ohio State University, University of
Pittsburgh, University of Portsmouth, Princeton University, the
United States Naval Observatory, and the University of Washington.

\end{document}